# CITATION CONTENT ANALYSIS (CCA): A FRAMEWORK FOR SYNTACTIC AND SEMANTIC ANALYSIS OF CITATION CONTENT

**Guo Zhang***

**Ying Ding**

**Staša Milojević**

School of Library and Information Science, Indiana University, 1320 E. 10th St,

Bloomington, IN 47405, USA.

E-mail: {guozhang, dingying, smilojev}@indiana.edu

* Corresponding author

**Abstract**

This paper proposes a new framework for Citation Content Analysis (CCA), for syntactic and semantic analysis of citation content that can be used to better analyze the rich socio-cultural context of research behavior. The framework could be considered the next generation of citation analysis. This paper briefly reviews the history and features of content analysis in traditional social sciences, and its previous application in Library and Information Science. Based on critical discussion of the theoretical necessity of a new method as well as the limits of citation analysis, the nature and purposes of CCA are discussed, and potential procedures to conduct CCA, including principles to identify the reference scope, a two-dimensional (citing and cited) and two-modular (syntactic and semantic modules) codebook, are provided and described. Future works and implications are also suggested.

**Introduction**

Scholarly impact is usually defined as the extent to which a researcher's work (e.g. a paper) has been used by other researchers (Bornmann et al., 2008). It has been considered an essential factor in estimating the value, credit and contribution of a certain paper, journal, institution or individual (Brown & Gardner, 1985). In citation analysis, this process is usually quantified as the citation counts provided by ISI Web of Knowledge, Scopus, Google Scholar, and so forth.

Traditionally, a citation is interpreted as an author A being influenced by the work of an author B, though without any attempt to specify the strength or direction of that influence. Additionally, it is assumed that each reference has made an equal contribution to the citing article. Therefore, in citation network analysis, citing behavior is usually simplified as a linear relationship—an edge in which node A cites node B—where nodes can be authors, papers, journals or institutions (Pinski & Narin, 1976). Based on this simplification, scholarly impact is also reduced to the number of citations. Obviously, the goal of traditional citation analysis is to answer two main questions: whether the two papers are connected through citations and how many citations a paper has accrued.

In fact, citing is an inherently complex behavior that is usually triggered by a variety of subjective factors (e. g., authors' intellectual and/or social motivations), and cannot be reduced to a simple linear relationship. Nicolaisen (2007), who has reviewed various theories of citing behavior and citation analysis, concluded that there is a widespread belief that citing behavior can be explained by evolutionary accounts of science and scholarship, and that it can be understood in terms of psychology, the normative theory and the social constructivist theory. Small (1978) suggested that citing constitutes an author's interpretation of the cited work, which is a process of meaning creating and symbol making. This process is considered as immaterial and sociologically orientated by Swales (1986). Thus, although by reducing citing to numbers and edges one can obtain a general and broad image of scholarly communication, this simplification risks ignoring the rich socio-cultural context of research.

The process of selecting citations is far from random in nature (Cronin, 1981; Small, 2011). It is driven by norms. For example Cronin (2004, p. 43) speaks of "the normative ghost in the machine" and Cronin (1984) and Small (1976) discuss procedural standards to which scientists typically adhere. Hereby we refer to such a set of norms as both individual norms and collective norms. The former explains citing behavior as triggered by individual motivations, while the latter suggests that citing can be learned in and shaped by specific groups or domains – i.e., how one cites is dependent on the discipline one belongs to (see for example Hellqvist, 2010, Milojević (under review)), or on certain characteristics of authors (their academic age, productivity and collaborative practices) regardless of their discipline or field (see for example Milojević (under review)).

With these concerns in mind, we propose Citation Content Analysis (CCA), as a promising addition to the traditional citation analysis methods that would enable syntactic and semantic, as well as quantitative and qualitative analysis of citation content. Traditional citation analysis is mainly quantitative (e.g. citation counts) and pays no attention to the actual context, while classical content analysis (CA) is essentially qualitative (e.g. codebook categories) and rarely applied to citation data. Endeavoring to make best of both methods and fill out the gap, CCA is adapted from CA but it is not a simple mixture of CA and citation analysis. Instead, CCA is mainly established on two rationales: 1) instead of being weighted equally citations should be granted different weights under different contexts; 2) qualitative measurements (e.g., how one cites) and quantitative measurements (e.g., number of citations) should be incorporated and mutually complementary.

Thus, CCA is conceptualized as an endeavor to describe the contextual relationship between citing and cited works, to indicate the social and intellectual interaction between different authors, to investigate the relative contribution of individual and collective norms to citing behavior, and to understand the nature and function of such behavior. In addition, with advances in computing capability and the growth of digital libraries and repositories, it is possible for CCA, as the next generation of citation analysis, to provide applicable classification schemes and to identify specific citing patterns across different domains, so as to facilitate further Natural Language Processing (NLP), and to develop scalable text-mining algorithms to extract associations hidden in large document collections.

In this paper, we briefly review the history and features of CA in traditional social sciences, and its previous application in Library and Information Science (LIS). Based on critical discussion of the limits of citation analysis, we propose that Citation Content

Analysis (CCA) should be the next generation of citation analysis that will improve traditional bibliometric research. Endeavoring to establish the theoretical framework of CCA, we discuss the nature and purposes of CCA. Potential procedures to conduct CCA are provided and described. Finally, future works and implications are suggested.

**A theory of citation: Why do we need a new method?**

It has become a convention for scientists and researchers to refer to earlier work (e.g. concepts, theories, methods, equipment, results, conclusions) that relates to, inspires or is used by their own work (Nicolaisen, 2007). Thus, citations have become intellectual linkages across academic and professional disciplines and can be used to study the nature and the development of different domains. In addition, citations can be studied from various perspectives, from information science (bibliometics) to linguistics (discourse analysis).

As early as 1986, Swales (1986) has pointed out one crucial problem in citation studies, that of existence of two relatively independent and separate orientations in citation analysis - quantitative description of bibliographical references, and qualitative interpretation of the symbolic indication embedded in citations. Information scientists usually focus on citation frequency but rarely take citers' rhetorical and linguistic choices into account, while linguists mainly focus on the embedded meanings in sample citations but fail to investigate the large-scale image by quantitative measures. According to Small and Klavans (2011), such a separation is mainly caused by data availability. Namely, quantitative researchers tend to use database that represent all sciences, but none of these databases provide full text data; while qualitative researchers tend to use relatively small and homogeneous data, gathering of which is labor-intensive and requires close reading, professional knowledge, and expert judgment (see also McCain and Turner, 1989).

Here (Table 1) we provide a brief summary of the conceptual origins, basic assumptions and popular methods of analysis of three main features of citation: (1) numerical, (2) literal, and (3) socio-cultural.

**Table 1. A theoretical foundation to understand the main features of citation**

| Features | Orientation | Conceptual origins | Basic assumptions | Analytical methods |
|---|---|---|---|---|
| Numerical | Explicit | The measurable nature of science | The number of citations can be considered impact indicators, or signs of breakthroughs. | e.g. Citation counts, citation frequency |
| Literal | Explicit/implicit | The symbolic nature of words | Words and linguistic choices in the citing text can indicate functions and sentiment of citations. | e.g. Discourse analysis, Natural Language Processing |
| Socio-cultural | Implicit | The individual and social nature of selections | The semantic content of citing contexts can suggest the citing motivations. | e.g. Content analysis, psychological experiments, surveys, interviews |

The most explicit feature of citation is numerical, which means that citations can be studied quantitatively. This view is closely tied to the idea of science of science, or the idea that we can apply the scientific methods to study the phenomenon of science itself. This approach has been widely used in the field of information science and has the counting the number of citations as its basis. The basic assumption here is that research impact is not intangible but measurable – in a quantitative way.

Citation is also literal. Citation is constructed by words (i.e., language), while language is a symbolic representation of concepts and ideas. Words can be used as linguistic cues to suggest a citer's intellectual process, cognitive interaction, attitude and sentiments. Words can also indicate whether the item in question is new, novel, or important, and thus be used to detect scientific breakthroughs, shifts, or revolutions (Small & Klavans, 2011). The literal feature of citation is neither completely explicit nor implicit. Words can be both explicitly quantified by parsing and counting, and carefully examined through a qualitative implicit process to determine their semantic meaning.

The third feature of citation is socio-cultural, which is implicit and difficult to obtain either from counting references or from the discourse analysis. The reason is that citation is a complex social system where both individual attributes and social dynamics interact and influence each other. Motivations behind every citation may vary greatly: personalized psychological process (e.g., Nicolaisen, 2007), citers' social environment and cultural background (e.g., Hjørland, 2000; 2002), normative tendency (e.g., Kaplan, 1965) governed by the internal norms in sciences proposed by Merton (1973), or an art of persuasion (e. g., Latour & Woolgar, 1986; White, 2004). There is no an existing method that provides a comprehensive analysis of all three features of citations.

Our framework for syntactic and semantic analysis of citation content draws from the existing theories of citing. We strongly believe that in order to make further advances in citation analysis two current orientations: quantitative description of bibliographical references and qualitative interpretation of citation context need to be combined. Therefore, we suggest a promising new approach (CCA) that incorporates content analysis (CA) and traditional citation analysis and is capable to comprehensively capture the nature of citation.

**Classical content analysis (CA): A flexible method**

As a classical research method, CA has been widely used and well defined in traditional social sciences. It was first used in Europe in the 17th century by the church to systematically examine content of early newspapers, then improved by sociologist Max Weber to study press coverage of political issues in Germany in 1910. From 1920s to 1950s, researchers started to develop the theoretical foundations for CA and applied it to mass communication (e.g. Berelson, 1952). Since 1960s, CA has been extended continuously and applied to other areas, e.g., anthropology, history, library and information studies (LIS), linguistics, management, political science, psychology, and sociology. In this process, researchers from different domains adapted CA to their unique research questions and goals. Thus, CA has become "a broadening of text aspects to include syntactic, syntagmatic, and pragmatic aspects of text, although not always within the same study" (White & March, 2006, p. 23). Not surprisingly, in today's digital era, CA is usually considered a flexible research method with the potential to incorporate both quantitative and qualitative approaches, conducted both manually and with computer assistance, which can be applied to many questions in different domains.

There are multiple definitions of CA reflecting its historical development and rich variants (e.g. conversational analysis, discourse analysis, ethnographic analysis, functional pragmatics, rhetorical analysis, and narrative semiotics. See Krippendorff, 2004), however, hereby we summarize it in terms of four characteristics: dynamics, resource, structure and operationalization.

*Dynamics: Systematic and objective*

Previous studies define CA as a systematic and objective research method. For example, Bauer (2000) identifies it as a systematic, replicable technique for coding data found in communication (of any type). Holsti (1969) suggests that CA is any technique for making inferences by objectively and systematically identifying specified characteristics of messages. Hereby "systematic" indicates at least two senses: systematic process of sampling of messages, and systematic structure (e.g. symbols-numbers, words, letters, computer codes, etc.) of sampled messages. "Objective" suggests that the analysis should make replicable, repeatable, and valid inferences from texts (or other meaningful matter) to the contexts of their use (Krippendorff, 2004). In this sense, CA is not a subjective interpretation of others' works but an incorporation of both quantitative and qualitative methods.

*Resource: Message-based*

In traditional social science research, CA is based on textual materials. Stone et al. (1966) propose that the ultimate goal of CA is to identify "specified characteristics within text" (p. 5), and to make specific inferences from text to other states or properties of its source. Therefore, CA was mainly used to systematically classify and count textual (word-based) units. However, in today's digital era the application of CA has been greatly expanded to diverse resources (e.g. images, videos, hyperlinks, etc.) besides pure texts. For example, Kress and van Leeuwen (1996), as well as Bell (2001), provide the framework of visual content analysis of images. Generally, the rich context and wide application of CA have led to its wide use in all symbolic data (messages in general).

*Structure: Syntax and semantics*

As a method embodying quantitative and qualitative components, CA focuses on both syntactic and semantic structures. The former refers to how symbolic data is organized and presented (e.g. feature/image/word frequencies, linguistic indicators, order of elements), while the latter demonstrates what is presented (meaning, both denotation and connotation), for example, themes, valuations and so forth. Syntactic and sematic structures are also called "analytical constructs, or rules of inference" (White & March, 2006, p. 27), which can be both quantified and qualified. It is based on these two analyzable structures that implicit meaning (i.e. content) that is embedded in the explicit presentation (messages) can be interpreted and understood.

*Operationalization: coding*

Constructing a systematic classification of message-based units is crucial for CA in which coding plays the central role. Cartwright (1953) even proposes that the terms *content analysis* and *coding* can be used interchangeably to emphasize the objective and systematic description of any symbolic behavior. According to White and March (2006), coding constitutes the body of CA and includes: 1) establishing coding scheme that allows for testing hypothesis, 2) coding data, 3) checking for reliability of coding, 4) adjusting coding process if necessary, 5) analyzing coded data, and 6) applying appropriate statistical test(s). Not surprisingly, for a number of reasons the most important step is establishing an appropriate coding scheme: 1) it is the coding scheme that operationalizes and qualifies the intangible concepts and implicit connotations; 2) valid and consistent assessment is achieved by establishing a coding scheme with clearly defined, comprehensive and mutually exclusive categories which represent relevant aspects (i.e., facets) of the data; 3) the reliability of research results and conclusions is highly correlated to the appropriateness of a coding scheme. Namely, the better a coding scheme is, the higher the interrater and intrarater reliability is (i.e., different coders will code the same item the same way or a single coder will code the same item the same way at different points in time (Krippendorff, 2004)). In general, constructing such an appropriate scheme is a complex and mainly qualitative process, which often involves careful, repetitive reading of the original messages, and modifying/re-modifying of the proposed scheme.

**Content analysis in Library and Information Science (LIS)**

Traditionally, CA has been used to determine authorship (from identifying personalized linguistic and rhetorical characteristics), examine patterns in documents, infer psychological or emotional states, and product evaluation. In library and information science (LIS) studies, CA has been extended to analyze different types of data (e.g. reference interviews, problem statements in published articles, job advertisements, etc.) in both qualitative and quantitative researches. For example, Pettigrew and McKechnie (2001) used a CA codebook (including three categories: Affiliation of First Author, Primary Subject of Article, and Type of Article) to analyze authors' use of theory in 1,160 articles that appeared in six information science (IS) journals from 1993–1998. In 2006, White and March (2006) provided a summary of selected examples of CA studies in LIS from 1991 to 2005, including identifying the reasons for selecting initial strategy in Web searches (White & Iivonen, 2001), developing a thesaurus of image-text relationships (Marsh & White, 2003), determining the nature of problem statements in LIS articles (Stansbury, 2002), and so forth.

In essence, the popularity of CA in LIS originates from its flexibility and appropriate match with the nature of LIS, which is shown in Table 2:

**Table 2. Advantages of CA in LIS**

|  | LIS features | Advantages of CA |
|---|---|---|
| **Main data type** | Raw data, existing historical data, archival records | Well suited to historical data and archival records |
| **Data amount** | Usually large amounts of data | Can deal with large amounts of data |
| **Procedure** | Replicable, retrievable, recordable | Offers a set of mature and well-documented procedures |
| **Cost** | Inexpensive, requires no contact with people | Inexpensive, requires no contact with people |
| **Boundary** | Flexible and interdisciplinary | Highly flexible, can be combined with other research methods (e.g. interviews, observation, statistics) |

As we mentioned in earlier section, there is one area of LIS where CA is still not widely used—citation analysis, as it is difficult to apply the qualitative essence of CA (e.g. codebook categories) to citation data. This is despite the fact that the idea of combining bibliometric methods with the full text analysis for the purposes of "context and content analysis of citations" (Cronin, 1984) was put forward and experimented with as early as 1960s (Glenisson et al., 2005). For example, in 1965 Lipetz identified 29 different reasons for citing and grouped them into four clusters. In the 1970s a number of researchers (e.g. Chubin & Moitra, 1975; Frost, 1979; Moravcsik & Murugesan, 1975; Oppenheim & Renn, 1978; Spiegel-Rosing, 1977) followed his ideas and proposed their own schemes to categorize and contextualize citations. Small (1982) and Cronin (1984) provided overviews of citation classification schemes. Some of the previous endeavors also focused on co-word analysis (e.g., Callon, Courtial, & Laville, 1991) or word analysis (e.g. Braam, Moed & Van Raan, 1991) in the context of evaluative bibliometrics to improve efficiency of co-citation clustering. However, these approaches are not actual CA.

In summary, difficulties of incorporating CA and citation analysis are of two kinds. First, citing behaviors are usually simplified as a linear one-dimensional relationship while CA is a descriptive and multi-dimensional method. Traditional citation analysis assumes that author A has been influenced by the work of author B, though without any attempt to specify the strength or nature of that influence. Additionally, it is assumed that each reference has made an equal contribution to the citing article. In contrast, CA endeavors to describe the citing behavior itself, as well as to interpret and understand the underlying motives for the observed pattern. Namely, it seeks to understand what the purposes, functions, attitudes, dispositions, and sentiments behind the citing behavior are and how these patterns are represented in citations to indicate authors' motivations. Second, it is difficult to establish an appropriate coding scheme for citing behaviors. As we discussed above, the most important step in CA is the creation of an appropriate coding scheme, which will establish a set of clearly defined, comprehensive and mutually exclusive categories. One reason for this difficulty lies in identifying sampling units, data collection units and units of analysis, which constitute the foundation of generating a coding scheme. According to White and March (2006), sampling units serve to identify the population and establish the basis for sampling. Data collection units are the units for measuring variables. Units of analysis are the basis for reporting analyses. In the context of citing behavior, to determine these units we need to make decisions regarding the following: should all scholarly work, or works in a given domain/discipline, be identified as sampling units? Should long papers, short papers, journal articles, conference papers, or books, be identified as data collection units for measuring?; and should a single sentence, or a cluster of sentences in which a reference is mentioned, be selected as units of analysis? Confusions generated from these questions indicate that the research domain (and its accepted writing pattern), the dominant genre(s), and the length/coverage of analytical units can influence the creation of a coding scheme, and can restrain the scope of its applications as well. In addition, citing behavior can be triggered by subjective factors. As Small (1978) suggests, citing constitutes an author's interpretation of the cited work, which is a process of meaning creating and symbol making. This process is considered as immaterial and too sociologically orientated by Swales (1986). Thus, it could be difficult to "re-interpret" authors' "interpretations" without deep background knowledge of the field and authors themselves.

Nevertheless, these complexities and challenges should not become the reason to avoid CA in citation analysis. Pioneering researchers from 1960s, 1970s to 1980s have done inspiring works in this area providing sound foundation and increasing "our understanding of the relationships which exist between citing and cited documents in the scientific literature" (Cronin, 1984). As followers of these pioneers, as well as with the aspiration to further investigate the norms and behaviors surrounding citations, we propose a framework for the new method Citation content analysis (CCA) that would introduce CA to the traditional citation analysis in a way that could revolutionize traditional bibliometric research. In the next sections, we discuss the nature and purposes of CCA, and propose potential procedures to conduct CCA.

**Citation content analysis (CCA): Nature and purposes**

Although the term *citation content analysis (CCA),* or similar terms, has been mentioned in several previous works (e.g. "content citation analysis", Swales, 1986; "citation content analysis", McCain & Turner, 1989), CCA in this paper introduces new implication and significance. Namely, CCA is not merely a text/word-based linguistic or discourse analysis approach. It is an endeavor to investigate all three features of citation: numerical, literal, and socio-cultural.

*A discourse approach for academic writing*

The main reason why CCA can become an appropriate method to analyze citing behavior comes from the nature of academic writing itself. It has been accepted and validated that CA is the most efficient when applied to semantically rich and logically consistent texts (e.g. Markoff et al., 1975). Academic writing meets all these requirements since it is formal, official, systematic and neutral to a great degree. Therefore, CCA is well suited for the analysis of texts of such a unique writing style.

In this sense, CCA can effectively organize, standardize and categorize both the explicit format and the implicit function of texts, so as to conduct systematic comparisons and reasonable interpretation. The coding procedure in CCA can divide, categorize and transform "this mass of documentation into an organized data file" (Markoff et al., 1975, p. 3) which is highly detailed and concrete. This process of operationalization can facilitate comprehending the intricate texts and promote the communication between different researchers to investigate the same data file, so as to shed light on the embedded motivation and connotations behind citations and citing behaviors. As Cronin (1981) points out, although such textual analysis cannot tell us all reasons why an author cites as he does, it may suggest very plausible reasons.

*A symbolic approach for conceptualizing citations*

Another reason why CCA can be and should be used to investigate citing behaviors is embedded in the symbolic nature of citations. Gilbert and Woolgar (1974) have distinguished citation from reference. Reference refers to the works mentioned in the reference section or bibliography of a journal article. A reference may be mentioned once or multiply in an article. Each mention is considered a citation. Thus, citations are the contexts in which references are made. According to Small (1978), citations can be considered to be "symbols of concepts or methods"–so citing is a process of creating cognitive links between concepts, procedures, types of data, and documents. This view also echoes Garfield's (1977) notion of cited documents as subject headings in an indexing system, and Gilbert's (1977) idea of citing as an author's device for persuading readers. As Cronin (1981) states, "citations are frozen footprints in the landscape of scholarly achievement; footprints which bear witness to the passage of ideas" (p. 16). The CCA can be used to operationalize and measure the intangible concepts and connotations, as well as the intellectual process of knowledge transferring and sharing. The proposed coding scheme and the analytical procedures can lead to a clearer and more specific image of interactions, conflicts, dialogs between different authors, documents, ideas, and paradigms, than traditional citation analysis.

*A macro-economic approach to citing behaviors*

In principle, CCA conceptualizes citing behavior as a decision making process in which citing is a way of information aggregation. In the perspective of macro economic theory, citing behavior is viewed as an incentive for the author to commit best effort to a task and make accurate predictions (Bacon et al., 2012), and a process of selection to reduce risks and optimize potential output. For example, what is the possibility of getting acceptance/acknowledgement if I choose to cite a work/author A? What if I cite B instead? What if I cite both A and B? What if I cite A and/or B in different ways? For individuals (authors), this is a prediction problem (e.g. Whether or not I will be benefited from citing this one?) and a selection problem (e.g. Which work I should cite to facilitate my success?) For collectives (a certain domain/field), this is an interaction problem between agents (personal motivations of members) and the community as a whole (established conventions of this field): What is the dynamic embedded in the "outcomes-based incentive system" (Bacon et al., 2012, p. 7) consisting of separate individuals? Using this approach, Othman and Sandholm (2010) have developed a single, deterministic decision rule: always select the action with the greatest probability of success, which has also been supported by Chen and Kash's (2011) study.

In this perspective, citing is not a random behavior or simply piling of all related works, especially considering the existing enormous literature corpus, the trend of interdisciplinary borrowing, and the regular limits of page numbers for publications. Instead, citing is a rational, selective, and comparative way to make best "economic" benefit. It is one way to decrease probability of failure but increase probability of success, decrease the risk (e.g. rejections, challenges, etc.) and cost (e.g. time, energy, social, cognitive, etc.) but increase security (e.g. acceptance, acknowledge, etc.) and output.

*An indicative approach to citing motivations*

Citing is a complex social and academic phenomenon that can be triggered by various subjective factors and cannot be reduced to linear relationship. Therefore, motivations behind citing behaviors, which are embedded in broad social contexts, cannot be interpreted merely through counting number of citations. Instead, CCA, with its theoretical and analytical roots in sociology and linguistics, and a grounding in actual discourse can provide a descriptive approach to indicate the in-depth citation motivations based on a broader context.

Different from the business and marketing activities, the "economic" benefit in this context is social rather than financial or monetary capital. Sociologists (e.g. Coleman, 1988; Portes, 1998; Putnam, 1995, 2000) have discussed the origins, definitions and applications of social capital, regarding it as a collective-based and intangible capital, which is generated by networks of relationships, reciprocity, trust, and social norms. Social capital facilitates both individual and collective action. Generally, social capital refers to the value of, and the economic (not monetary) benefits derived from the network of social relationships.

Based on this understanding, acquiring social capital can become an important motivation for citing and citation selections. Essentially, citing is a process of information aggregating to excess the personal limits of cognitive capabilities, endeavoring to break the boundary of individual rationality. Especially, in today's era of scholarly collaboration, scientific writing has become a dynamic process of borrowing, incorporating, creating and improving. Therefore, borrowing from others and self-creating based on previous works have been two crucial components of any scholarly work, as maintaining intellectual consistence and generating originality are equally important for any scientific researcher. As Chen and Kash (2011) state, "[i]nformation is often possessed by individual agents. Truthfully eliciting such information, resolving conflicting beliefs, and aggregating the dispersed information are key problems for achieving collective intelligence in multi-agent systems" (p. 1). Citing establishes the network of collaboration among different researchers, creates social capital in the forms of shared information, understanding, and knowledge, allows them to widen their horizons of understanding, increase their personal access to information and resources, achieve better outcomes, and, in turns, enhanced power (e.g. greater impact, higher reputation, and broader acknowledgement).

By means of citing, authors, as decision makers, both predict and influence their future impact. Similar to those decision makers in economic markets as discussed in Chen et al.'s (2011) work, authors can predict the effects of each of a set of possible actions – in this case, a set of all possible works an author can cite. This prediction is based on a process of cognitive evaluation in which authors pose questions such as: is this an appropriate work for me to cite?; can I incorporate this work into my work?; what kinds of benefits can I get from this work?; and how can I cite it to fulfill different purposes? After this reviewing, authors, as decision makers, can select an action to perform – citing A, or B, or both A and B.

**Potential procedure for CCA**

We have discussed complexities and challenges of incorporating both the descriptive, essentially qualitative CA method and the linear, simplified and mainly quantitative citation analysis method. Hereby we provide a framework for the potential procedure for CCA, endeavoring to optimize advantages of traditional CA and citation analysis, as well as to decrease their limitations.

**1.      Identify reference scope**

As we examined above, the fundamental challenge to create appropriate CCA method is to identify sampling units, data collection units and units of analysis, all of which constitute the premise to generate an applicable coding scheme. In other words, the main question is how to determine the reference scope. We propose three principles:

1)      **Principle of diversity** refers to the selection of heterogeneous sampling units, endeavoring to guarantee the generalization of the coding scheme. For example, one should use resources from different scientific domains (e.g., natural sciences, social sciences, humanities, etc.)

2）  **Principle of consistency** refers to the selection of  homogeneous data collection units, endeavoring to maintain the comparability of the coding scheme. For example, one should use the same genre (e.g., all conference papers, journal papers, or books.).

3)  **Principle of flexibility** refers to the flexible scope of units of analysis, depending on

syntactic or semantic categories in the coding scheme. One should use either single-sentence level or sentence-cluster level. At the single-sentence level, only the citing sentence that mentions previous work will be coded to identify the syntactic features of the citation (e. g., types of cited documents). At the sentence-cluster level, surrounding context (e. g., 1-2 sentences before or/and after the exact citing sentence) will be coded to indicate the semantic features of the citation (e.g., functions of citation). In this process, text-mining algorithms, Natural Language Processing (NLP), and topic modeling techniques can be used to determine and identify the scope of a cluster of sentences that are related to a given target reference.

**2.      Create the code book**

  The greatest challenge is to create an appropriate and applicable codebook for citations, which should be comprehensive but not too complicated, specific but not too trifle, be broadly applicable but not too general. Traditionally, a major criteria to evaluate social science research is its generalizability – the capability and reliability to make conclusions about the whole population based on results of the sample data, i.e., inference from the specific to the general (White & Marsh, 2006). However, when creating an appropriate and applicable coding scheme for CCA, the question becomes how to balance specificity and generalizability.

**Table 3. Summary of main coding schemes in citation analysis**

| Principles of coding | Sources | Example codes |
|---|---|---|
| **Type of motivation** | Lipetz (1965). | Group 4: Disposition of the scientific contribution of the cited paper to the citing paper (1-18 are included in Group 1-3 which are not |

| | | about type of motivation):<br>19.Noted only<br>20.Distinguished<br>21.Reviewed or compared<br>22. Applied<br>23.Improved or modified<br>24. Replaced;<br>25. Changed the precision (plus or minus)<br>26. Questioned<br>27. Affirmed<br>28. Refuted |
|---|---|---|
| **Level of importance** | Moravcsik & Murugesan (1975) | 1.Conceptual or operational<br>2.Organic or perfunctory<br>3.Evolutionary or juxtapositional<br>4.Confirmative/negational. |
| **Type of source** | McCain & Turner (1989), Frost (1979) | McCain & Turner (1989):<br>1. Research reports<br>2.Review articles<br>Frost (1979):<br>1.Primary sources<br>2.Secondary sources |
| **Function of citing** | Oppenheim & Renn (1978) and Spiegel-Rosing (1977) | 1. Methodological function (e.g., providing data, developing methods, etc.) 2. General function (e.g., historical background of a subject domain) |
| **Type of disposition/sentiment** | Frost (1987); Teufel et al, (2006); Small (2011) | Frost (1987): approval or disapproval;<br>Teufel et al, (2006): Weakness, contrast, positive, neutral<br>Small (2011): Importance, utility, report, consensus, uncertainty, differentiation, negation |
| **Location of mentioning** | Herlach (1978), Voos & Dagaev (1976) | Herlach (1978): Title/introduction, results/discussion, experimental<br>Voos & Dagaev (1976): Introduction, method, discussion, conclusion |

Table 3 is a summary of main coding schemes in citation analysis. The main problem with the existing coding schemes is their exhaustivity (i.e., researchers tend to create too many categories (more than 20) trying to capture all the possibilities). Such level of details tends to shrink the application scope of their schemes, and imposes enormous pressure on computer-assisted analysis.

In addition, although many LIS studies have been done to combine traditional bibliometric methods with full text analysis, and to develop classification schemes for citations, previous coding schemes tend to be subject to the following weaknesses: 1) Although addressing the same problem, the previous coding schemes tended to be disconnected from one another and to present different foci (Cronin, 1984). Some schemes focused on the function and quality of citations (e.g. Moravcsik & Murugesan, 1975; Oppenheim & Renn, 1978; Spiegel-Rosing, 1977), some emphasized the reasons for citing (e.g. Lipetz, 1965), and others paid attention to citation sentiment by identifying cue words (Teufel et al, 2006; Small, 2011). 2) The previous coding schemes were constructed more from the perspective of users' needs and perceptions, rather than from those of the citing authors, especially in terms of authors' citing motivations. Thus the contextual interrelations between the citing and cited works, and the distinctive features of the citing and cited authors are vague. Such an ambiguity poses difficulties on the explicit distinction between the citing and the cited, as well as on conducting an in-depth analysis of their interaction.

With a critical review of existing schemes, we propose a two-dimensional (citing and cited) and two-modular (syntactic and semantic modules) codebook for CCA. Based on grounded theory (Glaser & Strauss, 1967), the key approach we used to create our codebook is to learn from previous schemes and adapt to new challenges, to support both quantitative and qualitative measures, to distinguish citing-generated and cited-generated elements, to indicate both explicit and implicit principles, and to be inclusive of all the formats of resources in different domains. This codebook is also our endeavor to balance specificity and generalizability, as well as to investigate the interaction between individual norms (e.g. personal motivations) and collective norms (e.g. established regulations/conventions in a certain domain) in citing behavior. Table 4 summarizes this proposed codebook. In principle, this codebook is three-way orientated: attributes of the citing papers (e.g. category G, H, K, L), attributes of the cited papers (e.g. category A, B), and the attributes of the citing-cited interaction (e.g. category C, D, E, F, I, J).

**Table 4. Two-dimensional and two-modular code book for CCA**

| Orientation | Syntactic (Sy) | |
|---|---|---|
| | **Categories** | **Values** |
| **Cited** | A. Type of cited documents | 1.Journal article<br>2.Conference paper<br>3.Book/book chapter<br>4.Report/news<br>5.Link/personal blog<br>6. Others |
| | B. Type of authorship | 1.Single-authored<br>2.Multiple-authored |
| | C. Relation to the citing work | 1. Reciprocal (self-citatin)<br>2. Parallel (cite-coauthor)<br>3. Hierarchical (cite-author with high social capital) |
| | D. Location of mentioning | 1.Abstract<br>2.Introduction<br>3.Literature Review<br>4.Methodology<br>5.Results/discussion<br>6.Conclusion<br>7.Others (specify) |
| | E. Frequency of mentioning | 1. Once<br>2. 2 to 4 times<br>3. 5 times or more |

| | F. Style of mentioning | 1. Not specifically mentioning<br>2.Specifically mentioning but interpreting<br>3. Direct quotation |
|---|---|---|
| **Citing** | G. Type of citing documents | 1.Journal article<br>2.Conference paper<br>3.Book/book chapter<br>4.Report/news<br>5.Link/personal blog<br>6. Other |
| | H. Type of authorship | 1.Single-authored<br>2.Multiple-authored |

| **Orientation** | **Semantic (Se)** | |
|---|---|---|
| | **Categories** | **Values** |
| **Cited** | I. Function of citation | 1. Provide background information<br>2.Construct theoretical framework<br>3. Provide previous empirical/experimental evidence<br>4. Describe challenges and limits |
| | J. Disposition of citation | 1. Positive<br>2. Negative<br>3. Mixed<br>4. Neutral |
| **Citing** | K. Type of research domain | 1.Social sciences<br>2.Humanities<br>3.Natural sciences<br>4.Applied sciences and engineering |
| | L.  Type of research focus | 1. Theoretical research<br>2. Empirical research<br>3.Experimental research<br>4. Other (Specify) |

Depending on categories of the coding scheme, analysis can be conducted at either syntactic or semantic level. For the syntactic module, analysis is usually conducted at the single-sentence level, and can support the traditional bibliometric research on authorship, frequency, and all other quantitative measures.

Most importantly, Category C. (Relation to the citing work) can be used to study the latent connection between cited and citing works. This category is constructed based on theories of macroeconomics and social capital, implying the potential social triggers embedded in citing behaviors. It contains 3 categories: reciprocal, parallel and hierarchical. Reciprocal generally refers to self-citation, which suggests self-acknowledgement/development. Parallel refers to citing peers, co-authors, or collaborators, with a potential intention that the cited author(s) may cite back or reinforce possible collaborations in the future. For example, a sentence in one of Hjørland's works is: "Hjørland's (1991) criticized this approach in information science and began developing an alternative 'domain analysis' (Hjørland & Albrechtsen, 1995)". The two citations used in this sentence can be coded as "C1" (reciprocal) and "C2" (parallel) respectively, because the first citation is a self-citation and the second one cites a collaborative work. Hierarchical refers to citing prestigious authors with high social capital, potentially increasing one's own social or scholarly capital. To operationalize this coding, we propose that network analysis should be used to suggest a certain author's social capital based on betweenness, closeness, and degree centrality. By comparing the citing author's and cited author's social capital, one can decide whether C2 (parallel) or C3 (hierarchical) category should be assigned.

For a given cited work, identifying its location of mentioning (Category D) and counting its frequency of mentioning (Category E) in the same citing paper can suggest its level of significance, as well as the different citing patterns across disciplines. As Voos and Dagaev (1976) report, different disciplines exhibit different citation patterns in terms of the locations of citations. They argue that the contribution of a cited reference can be calculated based on the number of times it is cited and the location of those re-citations in the citing article and find that highly cited articles appear most often in the introduction In similar vein, Herlach (1978) maintains that if a work has been cited in the introduction or literature review section and is mentioned again in the methodology or discussion sections, it is likely that it makes a greater overall contribution to the citing paper than others that have been mentioned only once. Thus, it is possible that a reference that was mentioned more than 5 times in different sections of a paper is more important than a reference that was only mentioned once at the very end.

In addition, style of mentioning (Category F) can also indicate the importance of a cited paper. Bonzi (1982) argues that a reference that is cited by a paper but is not obviously mentioned in text can be considered less relevant than one that is discussed in depth within the text of the citing paper. Based on this understanding, we differentiate three styles of mentioning: not specifically mentioning, specifically mentioning but interpreting, and direct quotation. For example, utterance such as "Some studies have proposed…", "For example…" "e.g…" can be coded as "not specifically mentioning" (F1); utterance such as "Smith (2011) states that…" can be coded as "specifically mentioning but interpreting" (F2); utterance that contains "…. (Smith, 2011, P. xx)" can be coded as "direct quotation" (F3). It is possible that a reference with high frequency of type 3 mentioning is more important, or relevant, than a reference with low frequency of type 3 mentioning.

For the semantic module, analysis is usually conducted at the sentence-cluster level, so as to indicate the semantic features of the citation. We have created categories K (type of research domain) and L (type of research focus) only for citing and not cited papers in order to identify the potentially different citation patterns across various domains and papers with different research focuses, as well as to shed light on the embedded social context.

In terms of citing papers, let us take Paper A as an example. The paper is a critical review of historicist and pragmatic theories of information science concepts and was published in the *Journal of the American Society for Information Science and Technology* in 2009. We code Paper A as "K1: Social sciences" since it is published in an information science journal and not a philosophical research journal. It is also coded as "L1: Theoretical research", as it is an understanding and classification of theories of concepts in accordance with epistemological theories (empiricism, rationalism, historicism, and pragmatism). Thus its main contribution is theory-building. Papers that provide conceptual definitions, domain limitations, relationship-building, and predictions, offer framework for analysis, facilitate the efficient development of the field, and are needed for the applicability to practical real world problems fall into this category (Wacker, 1998).

Let us now look at Paper B. It describes the results of a content analysis of the Web sites of Fortune 100 companies and was published in *Corporate Communications: An International Journal* in 2000. We code paper B as "K1: Social sciences" since it is published in a business journal, which can be included in the general domain of social sciences. Paper B can also be coded as "L2. Empirical research", since it utilized both quantitative and qualitative methods to analyze the acquired empirical data (marketing communications differences across Fortune 100 websites), so as to test the proposed hypotheses and to answer the research questions. According to *the Oxford English Dictionary* (2nd Edition, 1989), empiric is derived from the ancient Greek for experience. Therefore, empirical data is based on direct or indirect observations and can be analyzed either quantitatively or qualitatively. Empirical research is any research that generates its findings on empirical data as its test of reality. Such research may also be conducted according to hypothetico-deductive procedures (Fisher, 1959), or Groot's (1961) empirical cycle (Observation- Induction- Deduction- Testing- Evaluation).

Paper C describes a systematic, unbiased, and comprehensive approach, termed "interactome capture", to define the mRNA interactome of proliferating human HeLa cells, published in *Cell*, 2012. It is obvious that Paper C should be coded as "K3: Natural sciences" since it is a research of biological cells and published in a biology journal. In addition, it should be also coded as "L3: Experimental research" not empirical research. Although experimental method is often misunderstood to be equivalent to empirical research, observational studies are not experiments. Experimental research is any research in which data are derived from the systematic manipulation of variables in an experiment (usually, laboratory experiment). Thus experimental research is more precise and rigid than empirical research in the sense that in an experiment the different "trials" are strictly manipulated so that an inference can be made as to causation of the observed change that results. In general, empirical research adopts a flexible "hypothetico-deductive" (Whewell, 1837) method while experimental research is constructed on rigid scientific tests and laboratory works.

In principle, publications in humanities are usually theoretical works; those in social sciences are often empirical works, while works in natural sciences and engineering are experimental researches. Although exceptions still exist, this principle of connecting "K. Type of research domains" and "L. Type of research focuses" can be used in computer-assist coding process.

Function of citation (Category I) is a major measurement to classify cited papers.

Instead of presenting too many details, we provide four values based on summarizing previous schemes and extracting the basic research flows: provide background information, construct theoretical framework, provide previous empirical/experimental evidence and describe challenges and limits. "Provide background information", which usually appears in introduction and literature review sections is generated from categories such as "background reading" and "historical" (Duncan et al., 1981), "general informational" and "historical" (Hodges, 1978), "historical background" and "description of other relevant work" (Oppenheim & Renn, 1978). "Construct theoretical framework", which mainly appears in the methods section, is summarized from categories such as "hypothesis or theory" and "calculation from theory" (Lipetz, 1965), "theory" and "development of ideas" (Duncan et al., 1981), "use of theoretical equation" and "use of methodology" (Oppenheim & Renn, 1978). "Provide previous empirical/experimental evidence" can appear in literature review, methodology and results/discussion sections, and is extracted from categories such as "evidential" (Hodges, 1978), "supplying information or data" (Oppenheim & Renn, 1978), "experimental details" (Duncan et al, 1981). "Describe challenges and limits", which usually appears in discussion and conclusion section, is developed from categories such as "questioned" and "refuted" (Lipetz, 1965), "disputing" and "criticism" (Duncan et al, 1981), "oppositional" (Hodges, 1978). These four values are demonstrated based on examples from different scientific domains such as theoretical, empirical and experimental research. Instances of "I1" (Provide background information) can be identified from the above examples of Paper A, Paper B and Paper C.

*Paper A: "Since philosopher of science Thomas Kuhn (1922–1996) wrote his famous book The Structure of Scientific Revolutions (1962), "paradigm" has been a popular term in many fields, although it has also been seriously criticized". (Theoretical-research)*

*Paper B: "The number of users of the Internet is estimated at 41 per cent of adults in the USA (Pew Research Center Survey, 1998)." (Empirical research)*

*Paper C: "Taking the natural variation between biological replicates into account, the bioconductor package DESeq (Anders and Huber, 2010) provides a statistical test for assessment of differential abundance of count data." (Experimental research)*

All these mentionings of previous works offer either historical background (e.g. Kuhn's theory) or information (e.g. facts of Internet use, statistical significance) regarding previous research as an explanation or elaboration of the author's research, no matter whether it is theory-focused, empirical-focused or experimental-focused works. Then examples of "I2" (Construct theoretical framework) are as follows:

*Paper A: "(e.g., "formal concept theory" by Priss, 2006). (Theoretical research)*

*Paper B: "Components of the marketing communications mix for Web sites include: advertising, sales promotions, public relations and direct marketing (adapted from Bennett, 1995). (Empirical research)*

*Paper C: "Our solution concept is the Perfect Bayesian Equilibrium (PBE) (Fudenberg and Tirole 1991)". (Experimental research)*

All the above citations cognitively represent symbolic concepts, i.e. specific terms/concepts (e.g. formal concept theory, components of marketing communication, PBE), which the citing author(s) can borrow, incorporate and develop to establish the principles and rationales of their own research. In addition, it is obvious that "I3" (Provide previous empirical/experimental evidence) rarely appears in theoretical research. For example, there is no such example in Paper A.

*Paper B: "Substantial empirical work has shown that prediction markets produce remarkably accurate forecasts (Berg et al. 2001; Wolfers and Zitzewitz 2004; Goel et al. 2010). (Empirical research)*

*Paper C: "It has been shown to interact with the 3' end stem loop of histone mRNA (Yang et al., 2006)." (Experimental research)*

Citations above refer to the empirical facts that support the citing author's work. Contextual cueing includes "substantial empirical work" and "it has been shown", which provide factual evidence or proof. Category "I4" (Describe challenges and limits) is closely related to Category J (Disposition of citation). Although there is an assumption that scientific writing tends to be objective and neutral, there is a distinction between "positive" (acknowledgement) and "negative" (questioning and challenging) citing. For example, citing for borrowing and establishing author's own research foundation, and citing for pinpointing the limits of previous research, can indicate author's sentimental tendency:

*Paper A: "However, the criticism of Kuhn's theory of paradigms suggests, among other things, that different 'paradigms' do not totally replace each other but exist together and compete with each other in all domains all the time (see, e.g., Mayr, 1997, pp. 98–99) (Theoretical research)*

*Paper B: "This is a common problem with other Web technologies in which user participation is necessary, for example, recommender systems (Raghavan, 2004)." (Empirical research)*

*Paper C: "Our model is not unique in suffering from a multiplicity of equilibria; multiple equilibria exist in many signaling games as well (e.g. Spence 1973). (Experimental research)*

To operationalize this coding, parsing and text mining can be used to identify the cue words such as "however" (Paper A), "but" (Paper A), "problem" (Paper B), "suffer" (Paper C), "nevertheless", "limit", "weak", "undermine", "ignore". Thus, all the above citations can be coded as "I2. Negative" based on these negative cue words. Such a vocabulary can be used for computer-assisted sentiment analysis.

Generally, the codebook we propose provides a relatively comprehensive and balanced framework to conduct CCA. Each citation in the text can be coded and assigned values using Categories A to I, covering dimensions of both citing and cited works, accounting for both individual and collective norms, as well as focusing on syntactic and semantic modules. The expected output is a comprehensive image of citations for different research purposes.

**Conclusion and future work**

Information science researchers have contributed to discussions of scholarly impact and have constructed a sophisticated and widely accepted method to measure it: citation analysis. For example, Voos and Dagaev (1976) suggested that the number of times a reference is cited in a paper provides some indication of its relevance to the citing paper's subject. However, Small (1987) also pointed out there is a great deal of evidence that influential papers are more highly cited than uninfluential ones, but there is no evidence that highly cited papers are highly influential. In other words, high number of citations is a necessary but not sufficient condition of "being influential".

In addition, citations do not exist in a vacuum but in an organized scholarly context that also reflects the rich socio-cultural properties, including motivations of citation, functions of citation, sentiments of citations, and so forth. Our goal is not a simplified, one-dimensional citation metrics, but an in-depth, multi-dimensional demonstration of the epistemological roles played by the citations in the citing paper, the heuristic values of the roles played by citations in the citing paper (Peritz, 1983), and the interactive network of social/scholarly capital implicated by citations in the citing paper. Based on this understanding, we have proposed a framework for the promising method -- Citation Content Analysis (CCA), to conduct syntactic and semantic analysis of citation content. We have also provided potential procedure for CCA, including principles to identify the reference scope, a two-dimensional (citing and cited) and two-modular (syntactic and semantic modules) codebook, and possible approaches to operationalize and apply this codebook via computer assistance.

Further work will concentrate on both facilitating current studies and inspiring future research trends. To expand current studies, we will test, modify and improve the proposed framework and apply CCA to a large-scale dataset, e.g. PubMed Central[1] data acquired from the U.S. National Institutes of Health's National Library of Medicine (NIH/NLM). This test will have both theoretical significance and applied importance.

Theoretically, it can shed light on a few epistemological questions in the current codebook. For example, syntactic features are usually hard to identify for citing papers, and thus only two categories (G and H) are provided in this paper. However, we hope that the analysis of the large-scale dataset with the special effort to understand the socio-contextual background of citations, can lead to identification of additional syntactic features. Another open question is that of how to balance deductive and inductive approaches. In this paper, we utilized an inductive approach to categorize citation motivation, while deductive approach should also prove to be quite useful to rationalize and model such categorization. For example, some researchers have already discussed the importance of a deductive approach (e.g. Börner et al., 2012) and applied it to citation studies (e.g. Chen, 2006; Chen & Hicks, 2004; Chen & Yu, 2000). By solving such questions, we hope that our framework will verify the appropriateness of incorporating quantitative and qualitative measures in citation analysis on a large scale. This can lead to a shift within citation analysis from the current purely numerical approaches to richer descriptive and contextual methods, which can provide more details than a simplified one-to-one relationship. Thus, this validation will enrich the current citation analysis and open a new frontier for computational linguistics that will focus on understanding and modeling citation patterns, which are quite different from natural language analyses.

Apart from the theoretical contribution, the empirical test will result in improved text mining and full text extraction algorithms, as well as advanced parsing and machine learning techniques. A possible output may be algorithms and software for intelligently processing language data.

Another possible venue of research is to combine CCA and topic modeling. In scholarly communication, topic modeling is important both at individual and collective level. Individually, it is a useful way to mine users' different opinions and attitudes toward various topics. Collectively, it can help analyze the heterogeneous academic domains and networks, facilitating community detecting. Thus, CCA and topic modeling are inter-dependent and mutually complementary.

In addition, authors' historical citing records can be organized by topic modeling and coded by CCA to map the change of opinions and sentiments these authors had regarding different topics through time, endeavoring to unveil their citing behavior patterns and to detect interrelations between citation motivation and topics. In this way, topic and content similarities can be used to predict authors' possible citing motivations and opinions on some specific topics, so as to visualize future citing patterns (e.g. stable, increase or decrease). There are at least three hypotheses that can be tested: (1) authors' opinion is tightly correlated with their topic preference, (2) authors' opinion is generally shaped in the context of a topic network and thus largely affected by direct influence from peers, and (3) the influence of authors' opinion can also be propagated through their indirect influence through topic network.

Using CCA together with topic modeling, means simultaneously incorporating topic factor and social-sentimental elements in a unified probabilistic framework, which would enable the analysis of the social opinion influence on scholarly networks, and the construction of an interactive influential network of the citing and the cited authors, works, and even patents (Tang et al., 2012a). This can be used to detect cross-domain collaborations, which exhibit very different patterns in terms of both content and topics (e.g. sparse connection, complementary expertise, and topic skewness) compared to traditional collaborations in the same domain (Tang et al., 2012b)

Concerning potential research trends in the future, we also suggest an emphasis on the new altmetrics (i.e. "the creation and study of new metrics based on the Social Web for analyzing, and informing scholarship", Priem et al., 2010) that would promote an awareness of the booming social media, and a disposition to interdisciplinary collaboration. With the boom of Web2.0, people, including scholars, became inclined to discuss, express and exchange ideas online. Priem et al. (2010) have pointed out that scholars are increasingly moving their everyday work to the web. For example, online reference managers Zotero and Mendeley each claim to store over 40 million articles (making them substantially larger than PubMed). The rise of social media, such as Facebook, Twitter and Microblog (as many as a third of scholars are on Twitter, and a growing number tend scholarly blogs), makes the expressions of scholarship and research impact more diverse than traditional citation metrics. This has led to new challenges to both citation analysis and CA in the field of LIS. Some of the traditional citation analysis methods are hard to apply to these new resources and at the very beginning classical CA was only developed for printed text. With a shared focus on the rich semantic data, CCA and altmetrics can provide potential solutions to the challenges generated by social media, shifting the focus to "how" and "why" from "how many". Future work is needed to correlate CCA, a new generation of citation analysis, and altmetrics, a new version of citation metrics, which will track impact both inside and outside the academia, impact of influential but unofficially cited work (e.g. Twitter mentioning, hashtags, Facebook @), and impact from sources that aren't peer-reviewed. This endeavor will balance new tools and existing measures, maintaining traditions while also adapting to new phenomena in the digital age.

In summary, we consider CCA a powerful yet feasible approach to improve the current citation analysis and a necessary supplement for traditional citation metrics. We are interested in understanding the impact of this new approach on analyzing the diversified forms of scholarly contribution in today's digital age, and its flexible use in interdisciplinary fields, which remain future work.

**Acknowledgement**

This manuscript is based upon work supported by the international funding initiative Digging into Data. Specifically, funding comes from the National Science Foundation in the United States (Grant No. 1208804), JISC in the United Kingdom, and the Social Sciences and Humanities Research Council of Canada. The authors would like to thank the other members of this grant for their comments on earlier presentations of the work.

1. PubMed Central® (PMC) is a free archive of biomedical and life sciences journal literature at the U.S. National Institutes of Health's National Library of Medicine (NIH/NLM). In keeping with NLM's legislative mandate to collect and preserve the biomedical literature, PMC serves as a digital counterpart to NLM's extensive print journal collection. Launched in February 2000, PMC was developed and is managed by NLM's National Center for Biotechnology Information (NCBI). From http://www.ncbi.nlm.nih.gov/pmc/about/intro/